\documentclass[12pt,letter]{article}
\usepackage{epsfig}

\newcommand{\beq}{\begin{equation}}
\newcommand{\eeq}{\end{equation}}
\newcommand{\bea}{\begin{eqnarray}}
\newcommand{\eea}{\end{eqnarray}}

\begin{document}

\title{{\bf Gluon distributions in nucleons and pions at a low resolution scale}}
\author{{\bf H.~R. Christiansen}\thanks{e-mail: hugo@cbpf.br}\\
{\normalsize {\it Centro Brasileiro de Pesquisas F\'{\i}sicas}, 
{\small CBPF - DCP}} \\  
{\normalsize {\it Rua Dr. Xavier Sigaud 150, 22290-180, Rio de Janeiro, Brazil}}\\
{\small and}\\ {\normalsize {\it Group of Theoretical Physics, Universidade 
Cat\'olica de Petr\'opolis}}\\
{\normalsize {\it Rua Bar\~ao de Amazonas 124, 25685-070, 
Petr\'opolis, Rio de Janeiro, Brazil}}\\ {\small and}\\
{\bf J. Magnin}\thanks{e-mail: jmagnin@uniandes.edu.co} \\
{\normalsize {\it Depto. de F\'{\i}sica, Universidad de los Andes,}}
\\ {\normalsize {\it AA 4976, Santaf\'e de Bogot\'a, Colombia }}
}

\maketitle
\date

\begin{abstract}
In this paper we study the gluon distribution functions in nucleons and pions 
at a low resolution $Q^2$ scale. This is an important issue since parton 
densities at low $Q^2$ have always been taken as an external input which is 
adjusted through DGLAP evolution to fit the experimental data at higher scales. 
Here, in the framework of a model recently developed, 
it is shown that the hypothetical cloud of {\it neutral} pions surrounding 
nucleons and pions appears to be responsible for the characteristic 
valence-like gluon distributions needed at the inital low scale. 
As an additional result, we get the remarkable 
prediction that neutral and charged pions have different 
intrinsic sea flavor contents.
\end{abstract}

\newpage

\section{Motivation}

Parton distribution functions (pdf) play an important role in particle 
physics as they describe the internal structure of hadrons in the framework 
of Quantum Chromodynamics. Pdf are also basic ingredients 
to calculate cross sections in hadron-hadron and lepton-hadron interactions.

Parton distribution functions have unique characteristics depending on each 
ha\-dron, which reflect the internal dynamics of the bound state. The pdf, 
$f_i(x,Q^2)$, are interpreted as the probability of finding a parton $i$ 
(quark or gluon) with a fraction $x$ of the hadron momentum when probed by 
the momentum transfer $Q^2$. The $Q^2$ dependence of pdf is succesfully 
described by the DGLAP~\cite{ap} evolution equations within perturbative 
QCD. However, in order to have adequate fits to Deep Inelastic Scattering 
(DIS) data, initial valence-like distributions at a low resolution scale must 
be considered even for sea quarks and gluons. This fact has been repeatedly 
noted by several authors (see e.g. Refs.~\cite{grv,cteq,mrs,suecos}) but
the origin of such initial distributions still remains unclear.

The origin of primordial low scale pdf should be traced back to the 
internal dynamics of the hadron 
bound state. Thus they must be related to the confining phase of QCD, which 
is ultimately the origin of hadrons as bound states of quarks and gluons. 
In this sense, the sea quark and gluon distributions at a low resolution scale 
can be related to the idea of the intrinsic sea of quarks and gluons proposed 
by Brodsky {\it et al.} at the beginning of the 80's~\cite{brodsky}. 
In their own 
words, ``intrinsic quarks and gluons exist over a time scale independent of 
any probe momentum, and are associated with the bound-state hadron dynamics''.
In contrast, the extrinsic sea of gluons and quarks have a purely perturbative 
origin, and their distributions show the characteristics of bremsstrahlung and 
pair production processes leading to the standard DGLAP perturbative QCD 
evolution.

The above considerations led us to draw a low $Q^2$ picture of 
hadrons in terms of effective quark degrees of freedom 
interacting among them through intrinsic sea quarks and 
gluons~\cite{christiansen}. The building blocks
are the so-called {\it valons} which are valence quarks appropriatedly 
dressed by their extrinsic sea~\cite{hwa2}.
Within this picture one can represent the hadron wave function 
as a superposition of hadron-like Fock states which we construct by means of a 
well-known recombination mechanism \cite{das-hwa}. 
For example, the proton wave function can be written as
\beq
\left|p\right> = a_0\left|p_{\circ}\right>+a_1\left|\hat{p}_{\circ}~g\right>
+\sum_i{a_i\left|B_i~M_i\right>}
\label{eq1}
\eeq
at some low $Q_v^2$ scale compatible with the valon picture of the proton. 
Here $\left|p_{\circ}\right>$ is a pure three valon state \cite{hwa2}, and the 
following terms are hadronic quantum fluctuations which emerge from
the (non-perturbative) interaction among valons. The role of these hadronic 
fluctuations is to dynamically generate the intrinsic sea of quarks and gluons, 
which is the necessary binding agent in order to have the hadron state. 
In this way, a consistent picture of the low $Q^2$ (input) scale of hadrons 
emerges in which hadrons are formed by constituent quarks plus intrinsic 
sea quarks and gluons. 

It is worth to stress  that in modern fits to DIS data, initial 
non-perturbative sea quark and gluon distributions are taken as an input 
which is adjusted by DGLAP evolution. In contrast, in our approach they 
are dynamically generated through the hadronic quantum fluctuations. Furthermore, 
the old fashioned fits to DIS data, in which all the hadron sea is perturbatively 
generated, are recovered by restricting the series of eq.~(\ref{eq1}) 
to the first term. 

This representation of the proton wave function led in a natural way 
to a $\bar{d}/\bar{u}$ asymmetry in the proton sea which closely 
describes~\cite{magnin} the most recent experimental data by the 
E866/NuSea Collaboration~\cite{e866}.
On the same footing, in Ref.~\cite{christiansen} a $s-\bar{s}$ asymmetry in the 
nucleon sea was calculated, qualitatively agreeing with the results 
of the last global analysis of DIS data~\cite{barone}.

In the following section of this paper, we determine the low $Q^2$
non-perturbative {\it gluon} distributions in nucleons, 
using the model introduced in Ref.~\cite{christiansen}.
In section~\ref{sec3} we explore the consequences 
of this picture on the non-perturbative structure of charged and neutral pions. 
Finally, section~\ref{sec4} is devoted to further discussion and conclusions.


\section{Intrinsic gluon distributions}\label{sec2}

To start with, let us consider a baryon at some low $Q^2_v$ 
scale. At this scale the baryon ground state is formed only by three 
valons~\cite{hwa2}. Quantum fluctuations will generate the non-perturbative 
$q\bar{q}$ sea. Following Ref.\cite{christiansen}, the non-perturbative sea 
has a two-step origin in our model. In the first step a valon emits a gluon which 
subsequently decays into a quark-antiquark pair. In the second, such quark and 
antiquark interact with  the valons giving rise to a bound $\left|MB\right>$ state.
Non-perturbative quark and antiquark distributions are then associated to the 
in-hadron meson and baryon valon densities.

The emission of a gluon out of a valon is a basic 
QCD process which can be adequately described in terms of the convolution 
of the valon distribution, $v(z)$, with the Altarelli-Parisi 
$P_{gq}(z)$ and $P_{qg}(z)$ splitting functions~\cite{ap}. 
In this way, the quark and antiquark initial distributions 
are given by~\cite{christiansen}
\beq
q(x) = \bar{q}(x) = N \frac{\alpha_{st}^2(Q^2_v)}{(2\pi)^2}
\int_x^1 {\frac{dy}{y} P_{qg}\left(\frac{x}{y}\right) \int_y^1
{\frac{dz}{z} P_{gq}\left(\frac{y}{z}\right) v(z)}} \; .
\label{eq2}
\eeq  
Notice that, as the valon distribution does not depend on $Q^2$, the scale dependence 
in eq.~(\ref{eq2}) only arises through the strong coupling constant $\alpha_{st}$. 
At this stage the scale is fixed to the valon scale, $Q^2_v$, which is tipically 
about $1$ GeV$^2$. Indeed, in Ref.~\cite{hwa2} 
the valon scale was estimated to be of the order of $Q_v^2\simeq 0.64$ GeV$^2$ for nucleons. 
Consequently, pair creation can be 
safely evaluated in a perturbative way since $(\alpha_{st}/2\pi)^2$ is still 
sufficiently small.

The perturbative $v\rightarrow g\rightarrow q\bar{q}$ process is the source of both 
the extrinsic and intrinsic seas. The difference among them rest on the fact that an 
intrinsic $q\bar q$ pair interacts with the remaining valons while an 
extrinsic $q\bar q$ pair not. In this sense, the extrinsic sea, which is 
purely perturbative, forms the structure of valons~\cite{hwa2}.

The second step involves the interaction of such $q\bar{q}$ pair with valons thus 
giving rise to the $\left|MB\right>$ bound state. As they are in the realm of 
confinement, the interactions of the $q\bar q$ pair with valons must be evaluated 
by means of effective methods. Notice also that, for such interactions take place, 
the initial (perturbative) $q\bar q$ pair must be sufficiently longlived. Since the 
characteristic lifetime of such a perturbative $q\bar q$ pair scales as $1/m_q$, 
light and strange quarks should be largely available to interact with valons thus 
producing the $\left|MB\right>$ hadronic quantum fluctuations.

Then, assuming that the in-hadron meson and baryon 
formation arises from mechanisms similar to those at work in the 
production of real hadrons, we can evaluate the in-hadron meson probability 
density by using the Das-Hwa recombination approach~\cite{das-hwa}.

In the recombination model, the probability density for the production of 
a real meson as a function of its fractional momentum is given by 
the convolution of a two-quark distribution with a suitable 
recombination function. The two-quark distribution is given in terms  of 
the single-quark distributions of the initial hadron which will be the 
valence quarks in the final meson. The recombination function is chosen in 
such a way that it favors the recombination of quarks with similar 
momentum fractions. 

Thus, in our model, the in-hadron meson distributions  
are given by 
\beq
P_{M_iB_i}(x) =  \int_0^1 \frac{dy}{y} {\int_0^1\frac{dz}{z} F_i(y,z) 
R(x,y,z)}\; ,
\label{eq3}
\eeq
where
\beq
R(x,y,z) = \alpha\, \frac{yz}{x^2}\, \delta \left(1 - \frac{y+z}{x}\right) 
\label{eq4}\\
\eeq
is the recombination function~\cite{das-hwa}, and 
\beq
F_i(y,z) = \beta\, yv(y)\, z\bar{q}_i(z)\, (1-y-z)^a
\label{eq5}
\eeq
is the valon-antiquark distribution~\cite{magnin}. In eqs.~(\ref{eq3})-(\ref{eq5}), 
$x$, $y$ and $z$ are the momentum fraction of the in-hadron meson, 
the valon and the antiquark respectively. The index $i$ runs over 
different quark flavors, depending on the meson being formed. 

Due to momentum conservation, the in-hadron meson and baryon probability 
densities are not independent but correlated by
\beq
P_{M_i B_i}(x) = P_{B_i M_i}(1-x) \; ,
\label{eq6}
\eeq
with an additional correlation in velocity given by
\beq
\frac{\left< xP_{M_i B_i}(x)\right>}{m_{M_i}} = 
\frac{\left< xP_{B_i M_i}(x)\right>}{m_{B_i}} \; .
\label{eq7}
\eeq
The above constraint, eq.~(\ref{eq7}), which is needed in order to build a 
$\left|M_iB_i\right>$ {\it bound} state, fixes the exponent $a$ in 
eq.~(\ref{eq5})~\cite{magnin}.

The hadronic fluctuations so far described can be interpreted 
as the origin of the intrinsic quark-antiquark sea. As a consequence, since the 
resulting $q$ and $\bar q$ sea distributions belong to different hadronic states 
in the $\left|M_iB_i\right>$ fluctuation, intrinsic quark and antiquark probability 
densities in baryons are unequal in a general way.

At this point, a judicious analysis of what fluctuations should be included 
in an expansion like eq.~(\ref{eq1}) must be made.
For definitness, consider the proton wave function. Taking into account 
mass values and quantum numbers, the main fluctuations of the 
proton should be the $\left|\pi^+\,n\right>$, $\left|\pi^+\,\Delta^0\right>$ and 
$\left|\pi^-\,\Delta^{++}\right>$ virtual states, with probabilities 
$|a_{\pi\,n}|^2$ and $|a_{\pi\Delta}|^2$ respectively. Differences between the 
$\left|\pi^+\,\Delta^0\right>$ and $\left|\pi^-\,\Delta^{++}\right>$ probabilities 
are taken into account by Clebsh-Gordan coefficients which ensure the correct 
global isospin of the fluctuation. Thus, we obtain $\frac{1}{6}|a_{\pi\Delta}|^2$ 
and $\frac{1}{2}|a_{\pi\Delta}|^2$ for $\left|\pi^+\,\Delta^0\right>$ and 
$\left|\pi^-\,\Delta^{++}\right>$ respectively. On the other hand, the probability 
of the $\left|\pi^+\,n\right>$ bound-state is 
$\frac{2}{3}|a_{\pi\,n}|^2$. 
The coefficients $|a_{\pi\,n}|^2$ and $|a_{\pi\Delta}|^2$ are given by 
$(N\alpha\beta)_{\pi\,n}$ and $(N\alpha\beta)_{\pi\Delta}$ respectively. Their
numerical values result from comparison with experimental data.
Fluctuations like $\left|\rho\,N\right>$, $\left|\rho\Delta\right>$, etc. which 
could contribute, for instance, to the $\bar{d}-\bar{u}$ asymmetry in the 
proton, are far off-shell~\footnote{Note that these are even more suppressed 
than strange, $\left|K\,H\right>$, fluctuations.} and can be safely neglected.

Remarkably, as shown in Ref.~\cite{magnin}, this scheme 
leads to a $\bar{d}/\bar{u}$ asymmetry in the proton which closely 
describes the experimental data by the E866 Collaboration~\cite{e866}. 
Including fluctuations to Kaon-Hyperon states, 
$\left|K\,H\right>$, a $s-\bar{s}$ asymmetry in the proton sea 
arises~\cite{christiansen} which qualitatively agrees with results from a 
recent global analysis of DIS data~\cite{barone}.

\begin{figure}[htb] 
\centerline{\psfig{figure=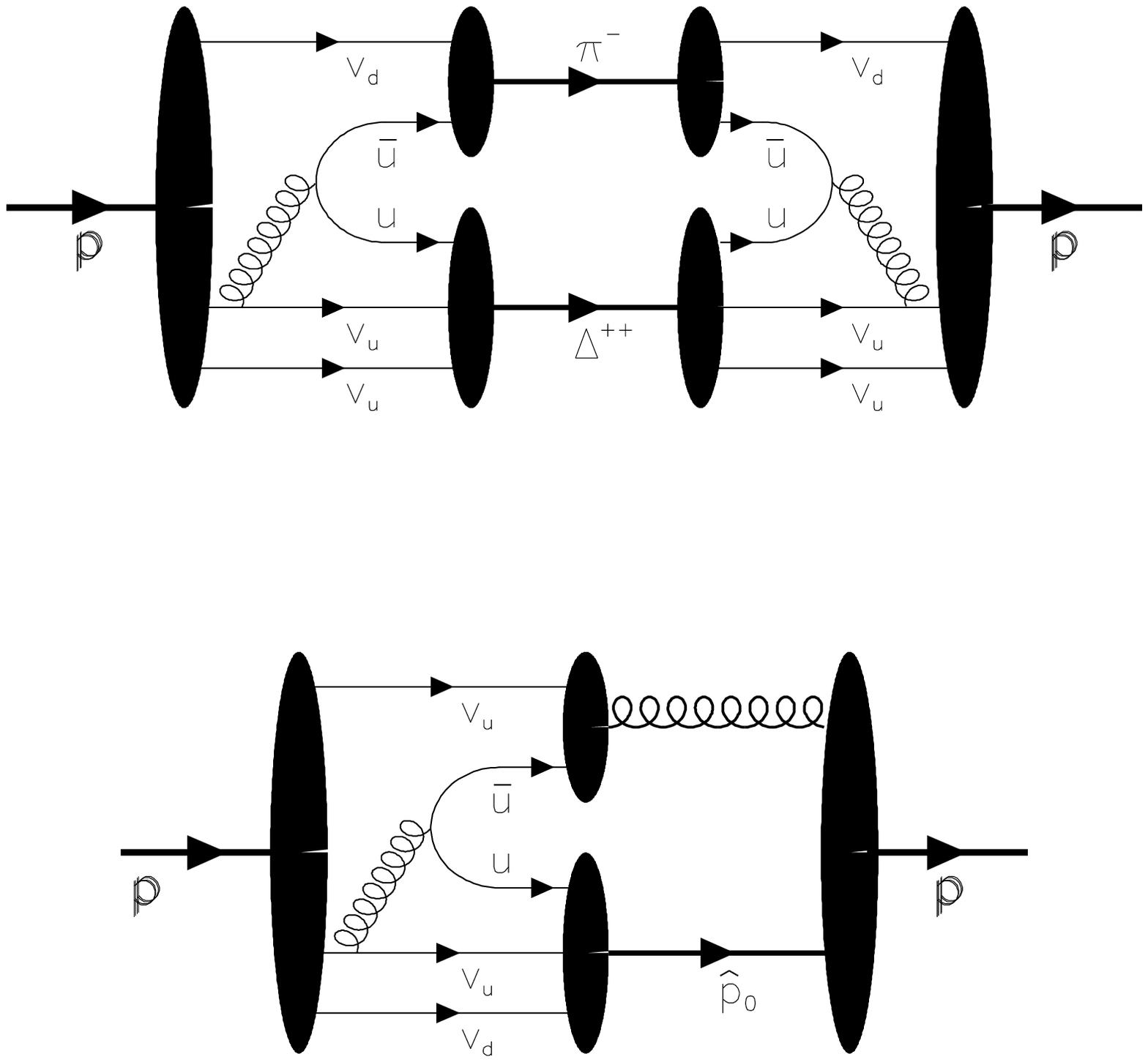,height=5.5in}}
\caption {Diagramatic representation of the process giving rise to a 
$\left|\pi^-\Delta^{++}\right>$ fluctuation of the proton (upper). 
Process leading to the generation of intrinsic gluons through the 
recombination of a $v_q~\bar{q}$ pair into a gluon (lower).}
\label{fig1}
\end{figure}

Fluctuations of the proton to $\left|\pi^0\,p\right>$ and 
$\left|\pi^0\,\Delta^+\right>$ states do not contribute to the intrinsic 
quark and antiquark structure. The reason is that the formation of 
a $\pi^0$ in-proton state must be inhibited due to its neutral flavor structure 
$u\bar{u}-d\bar{d}$. This happens because $v_q\bar{q}$ objects can 
annihilate rapidly into a gluon while $v_q\bar{q}\,'$ cannot ($q'$ is a 
quark of different flavor to the $q$ one. See Fig.~\ref{fig1}). 
Notice also that an unflavored object like a 
$v_q\,\bar{q}$ pair has itself the quantum numbers of a gluon. Thus, a hypothetical 
$\left|\pi^0\,p\right>$ fluctuation 
does not contribute to the sum over $\left|M_iB_i\right>$ in 
the RHS of eq.~(\ref{eq1}) but to the second term, $\left|\hat{p}_\circ~g\right>$, 
providing a source of valence-like gluons in the proton. The proton-like object 
accompanying the non-perturbative gluon in the $\left|\hat{p}_\circ~g\right>$ 
fluctuation must have the same flavor structure than the $p_\circ$ in the 
$\left|p_\circ\right>$ state. However, as far as the gluon is in a color octect 
state, the $\hat{p}_\circ$ must be colored. It is worth noting that, on general 
grounds, hadrons in a $\left|MB\right>$ fluctuation must be colored. However, they 
can be identified with real hadrons regarding other quantum numbers 
like as flavor, isospin, etc.

The time scale over which both the $\left|\pi^+n\right>$ and the 
$\left|\hat{p}_\circ~g\right>$ fluctuations exist should be approximately 
the same. In fact, the characteristic lifetime of a $\left|M\,B\right>$ fluctuation 
is proportional to $1/\Delta E$, where $\Delta E$ is the energy difference 
between the $\left|M\,B\right>$ and the proton states in an infinite momentum 
frame. Thus, for a generic $\left|M\,B\right>$ state we have
\beq
\tau_{\left|M\,B\right>} \sim \frac{1}{\Delta E} = \frac{2P}{
\left[\frac{\hat{m}_M^2}{x_M} + \frac{\hat{m}_B^2}{x_B} - m_p^2\right]} \; ,
\label{eq8a}
\eeq
where $P$ is the momentum of the proton in the infinite momentum frame, $m_p$ is 
the proton mass,  and $x_M$ and $x_B$ are the momentum fractions carried by the 
meson and baryon in the fluctuation. $\hat{m}_{M,B}^2=m_{M,B}^2+k_T^2$  is the 
transverse masses squared of virtual hadrons in the fluctuation. Given the smallness 
of the pion mass, we can assume that in-nucleon pions and non-perturbative gluons 
have similar transverse masses, then the characteristic lifetimes of the 
$\left|\pi^+~n\right>$ and $\left|\hat{p}_\circ~g\right>$ fluctuations 
must be approximately the same.

In this approach, the shape of the non-perturbative gluon coming from the 
$v_{q}\,\bar{q}$ pairing above described can be estimated by using the 
recombination model. Actually, as the origin of the non-perturbative gluon 
is the recombination of a valon with an antiquark of the same flavor, the 
momentum distribution of intrinsic gluons in the $\left|\hat{p}_\circ~g\right>$ 
fluctuation is simply given by
\bea
{g}^{NP}(x,Q_v^2) & = & |a_1|^2~P_{pg}(x,Q_v^2) \nonumber \\
& \equiv & |a_1|^2 \frac{(1-x)^{12.9}}{x}\int_0^x{dy\,y\bar{q}(y)\,
(y-x)v_{Nq}(y-x)} \; ,
\label{eq8}
\eea
where $\bar{q}(x)$ is given by eq.~(\ref{eq2}) and $v_{Nq}(x)$ is the distribution 
of the $q$-flavored valon in the nucleon given by~\cite{hwa2}
\beq
v_{Nq}(x) = \frac{105}{32}\sqrt{x}(1-x)^2 \; .
\label{eq9}
\eeq
Coefficients $N$, $\alpha$, and $\beta$; coming from the definition of the initial 
perturbative $\bar q$, the recombination function and the valon-antiquark distribution 
in eqs.~(\ref{eq2}), (\ref{eq4}) and (\ref{eq5}) respectively; were included in the 
definition of  $|a_1|^2$, the probability of the fluctuation. The intrinsic gluon 
probability distribution, $P_{pg}(x,Q_v^2)$, is accordingly normalized to unity.

The intrinsic gluon probability density in the $\left|\Delta^+~g\right>$ fluctuation 
can be estimated on similar basis. However, this fluctuation should be suppressed 
with respect to the $\left|\hat{p}_\circ~g\right>$ one.

\begin{figure}[htb] 
\centerline{\psfig{figure=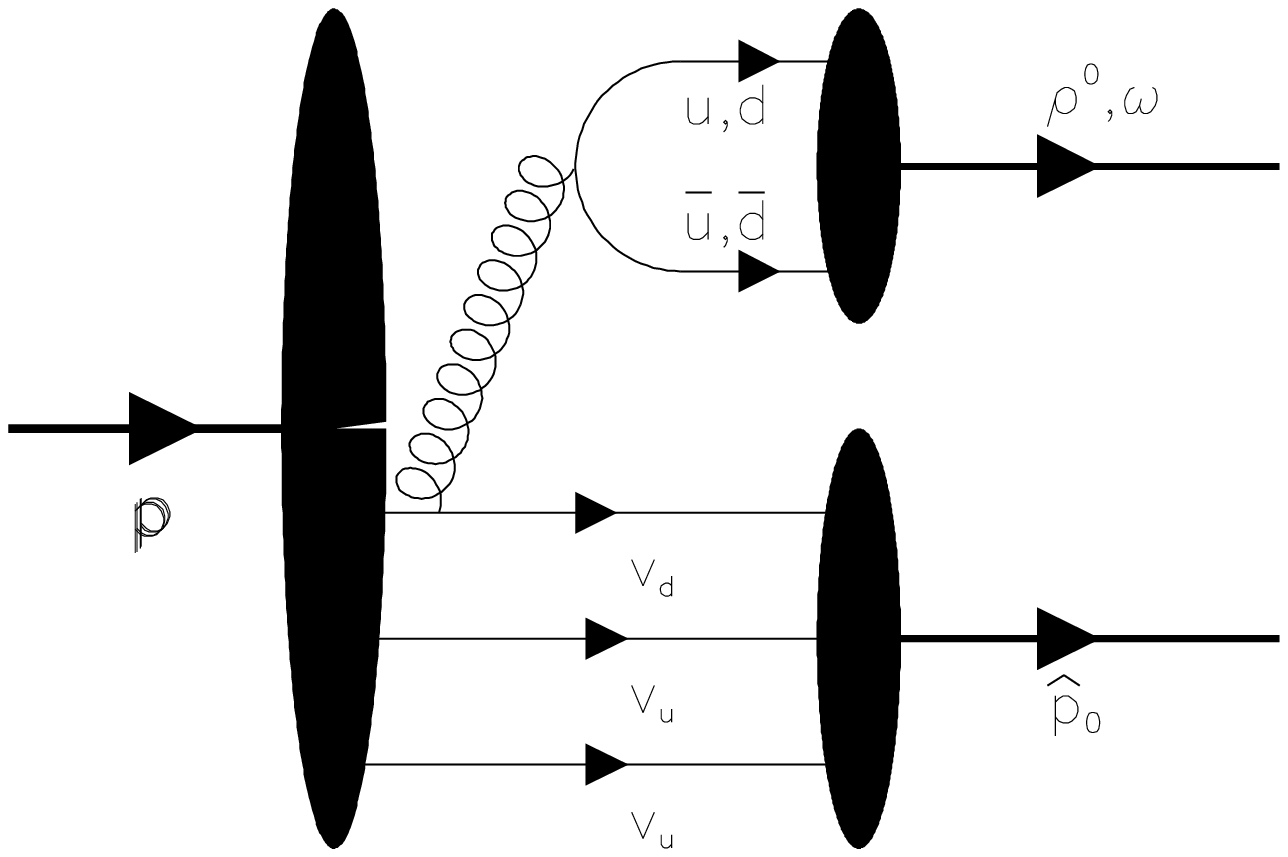,height=5.5in}}
\caption {OZI rule suppressed $\left|\rho^0(\omega)\,\hat{p}_\circ\right>$ fluctuation.}
\label{fig1a}
\end{figure}

Another way to have a proton fluctuation containing a proton-like object
and an unflavored neutral meson would be through the self-recombination of 
the $q-\bar{q}$ pair produced by the gluon splitting of eq.~(\ref{eq2}) 
(see Fig.~\ref{fig1a}). In this case, the unflavored neutral meson must be a vector 
meson like a $\rho^0$ or $\omega$. This is necessary to preserve the vectorial 
character of the initial perturbative gluon. However, this kind of fluctuations, consisting 
of a disconnected $\rho^0$ or $\omega$ and a $\hat{p}_\circ$, are strongly 
suppressed by the OZI rule.

\begin{figure}[htb] 
\centerline{\psfig{figure=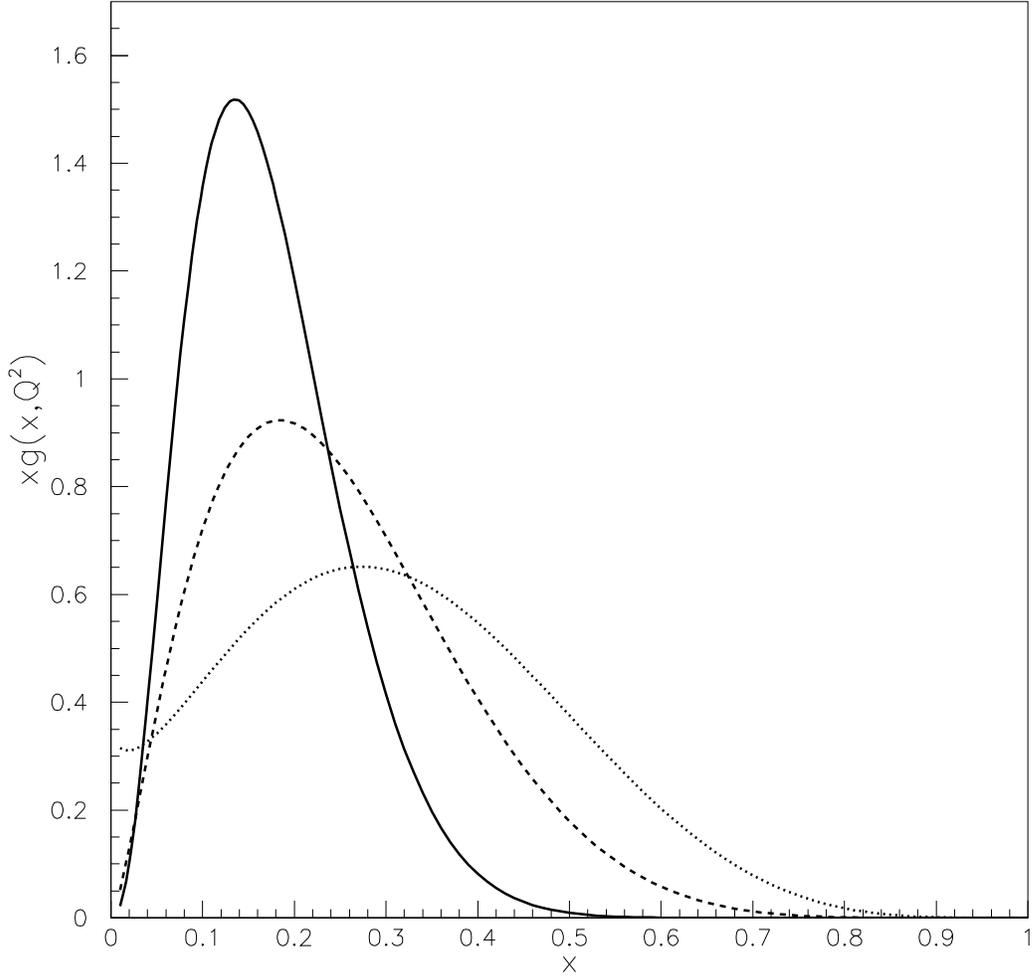,height=6.0in}}
\caption {Intrinsic gluon distribution for {\it nucleons} predicted by the model at 
$Q_v^2 \simeq 0.64$ GeV$^2$ (solid line) compared to the valence gluon distribution given 
by the model of Ref.~\cite{suecos} at $Q_0^2 \sim 1$ GeV$^2$ (dashed line) and the 
initial GRV-94 HO gluon distribution~\cite{grv} at $Q^2_0 = 0.4$ GeV$^2$ (point line). 
Model curves were normalized to the value of the integral over $x$ of the GRV-HO gluon 
distribution.}
\label{fig2}
\end{figure}

In Fig.~\ref{fig2}, the intrinsic gluon distribution at the valon scale 
given by eq.~(\ref{eq8}) is compared to the initial GRV-94 HO~\cite{grv} 
gluon distribution and the valence gluon distribution calculated in a Monte 
Carlo based model of the proton~\cite{suecos}
\footnote{In \cite{suecos}, a model for hadrons is proposed in which primordial 
pdf corresponding to valence quarks and gluons are assumed to have Gaussian 
distributions with widths fixed from experimental data. These initial pdf 
are then complemented with contributions coming from $\left|MB\right>$ fluctuations. 
In this model, intrinsic gluons are supposed to be present from the very 
beginning. In our model we are proposing a dynamical mechanism for their generation. 
This is the main difference between these two approaches.}.


\section{Non-perturbative structure of pions}\label{sec3}

Similar mechanisms should be at work in other physical hadrons, like pions. 
Indeed, if we expand the pion wave function as
\beq
\left|\pi^{\pm,0}\right> = b_0\left|\pi_{\circ}^{\pm,0}\right> + 
b_1\left|\hat{\pi}_{\circ}^{\pm,0} g\right> + 
\sum_i b_i {\left|M_iM_i'\right>} \; ,
\label{eq10}
\eeq
we can identify the would be $\left|\hat{\pi}_{\circ}^{\pm,0}\pi^0\right>$ fluctuation 
with the $\left|\hat{\pi}_{\circ}^{\pm,0}g\right>$ 
one, as we made for nucleons. In this way, the intrinsic gluon contribution to 
the pion low $Q^2$ structure is given by a similar expression to that of 
eq.~(\ref{eq8}), 
\bea
{g}_{\pi}^{NP}(x,Q_v^2) & = & |b_1|^2~P_{\pi g}(x,Q_v^2) \nonumber \\
& = & |b_1|^2~\frac{(1-x)}{x}\int_0^x{dy\,y\bar{q}_{\pi}(y)\,(y-x)v_{\Pi}(y-x)} \; ,
\label{eq11}
\eea
where $\bar{q}_{\pi}(x)$ is an antiquark distribution analogous to $\bar{q}(x)$ 
in eq.~(\ref{eq2}) but with a valon distribution for pions $v_{\Pi}(x)=1$, as 
given in \cite{hwa2}. Regarding the exponent related to velocity correlation, 
it turns out to be simply $a=1$. For the valon scale in pions we have used the 
same value than for nucleons, $Q_v^2\simeq 0.64$ GeV$^2$. 

As explained above, a $v_q\,\bar{q}$ pair should recombine into a gluon instead 
of forming a neutral pion structure. Thus the same conclusions drawn for 
nucleons hold for fluctuations of the pion into a Fock state containing the initial 
pion plus a neutral pion-like object.

\begin{figure}[htb] 
\psfig{figure=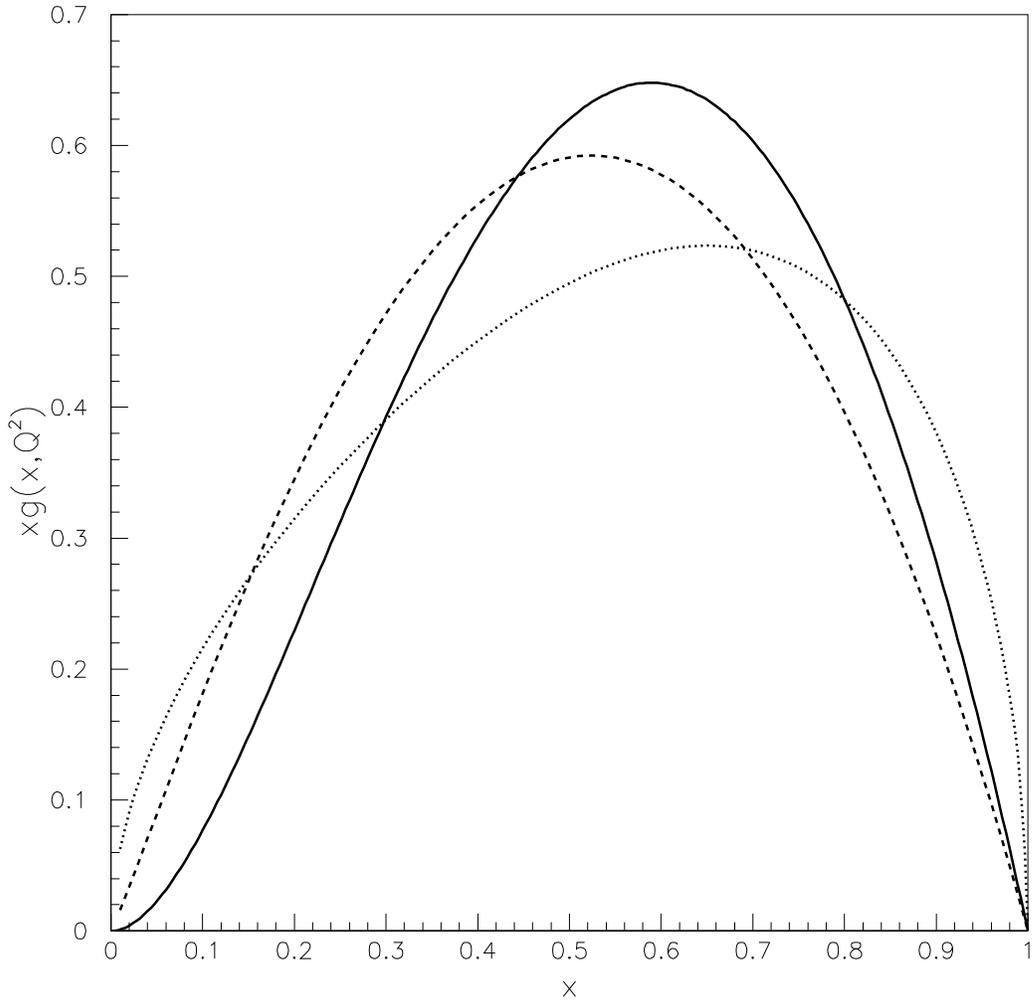,height=6.0in} 
\caption {Intrinsic gluon distribution for {\it pions} predicted by the model 
at $Q_v^2 \simeq 0.64$ GeV$^2$ (solid line) compared to the valence gluon distribution 
given by the model of Ref.~\cite{suecos} at some hadronic low $Q^2_0$ scale 
(dashed line) and the initial GRV-P HO gluon distribution~\cite{grv-pion} at 
$Q^2_0 = 0.3$ GeV$^2$ (point line). 
Model curves were normalized to the value of the integral over $x$ of the GRV-P HO gluon 
distribution.}
\label{fig3}
\end{figure}

In Fig.~\ref{fig3}, the ${g}_{\pi}^{NP}(x,Q_v^2)$ predicted by the model 
is shown, and compared to the GRV-P HO~\cite{grv-pion} distribution 
and the gluon distribution in pions obtained in Ref.~\cite{suecos}.
It is interesting to note that our model predicts intrinsic gluon distributions 
carrying more average momentum in pions than in nucleons. 
This is a consequence of the fact that the $v_q\,\bar{q}$ pair giving rise to 
the intrinsic gluon carries more average momentum in a 
$\left|\hat{\pi}_\circ^{\pm,0}\,g\right>$ fluctuation than in a 
$\left|\hat{p}_\circ~g\right>$ fluctuation.

The fact that pions do not fluctuate into $\left|\hat{\pi}_\circ^{\pm,0}\pi^0\right>$ 
states but into $\left|\hat{\pi}_\circ^{\pm,0}\,g\right>$ ones has additional consequences on 
their low $Q^2$ scale structure. For example, for {\it charged} pions 
the first contribution 
to the sum in the RHS of eq.~(\ref{eq10}) arises from the $\left|K^+\bar{K}^0\right>$ 
($\left|K^-K^0\right>$) fluctuation of the $\left|\pi^+_\circ\right>$ 
($\left|\pi^-_\circ\right>$) state. Thus,
\bea
\left|\pi^+\right> &=& b_0\left|\pi^+_{\circ}\right> + b_1 \left|\pi^+_{\circ}\,g\right> 
+ b_3 \left|K^+\,\bar{K}^0\right> + ... \; \\
\left|\pi^-\right> &=& b_0\left|\pi^-_{\circ}\right> + b_1 \left|\pi^-_{\circ}\,g\right> 
+ b_3 \left|K^-\,K^0\right> + ... \; .
\label{eq14a}
\eea 
Then the first contribution to the intrinsic $q\,\bar{q}$ sea
arises in the strange sector and there are no $u\bar{u}$ and $d\bar{d}$ 
intrinsic seas in charged pions. The structure of charged pions at the low $Q_v^2$ 
scale is thus given by
\bea
v_{q/\pi^\pm}(x,Q_v^2) & = &
|b_0|^2\,v_{\Pi}(x) 
+ |b_1|^2\,\int_x^1{ \frac{dy}{y}P_{g\pi}(y)\,v_\Pi\left( \frac{x}{y} \right)}
\nonumber \\ && + |b_3|^2\,\int_x^1{ \frac{dy}{y}P_{KK}(y)\,v_{Kq}
\left( \frac{x}{y} \right) } \nonumber \\
s_s(x,Q_v^2) & = & \bar{s}_s(x,Q_v^2) =  |b_3|^2
\int_x^1{ \frac{dy}{y} P_{KK}(y)\,v_{Ks} \left( \frac{x}{y} \right) } ,
\label{eq14}
\eea
where $v_{q/\pi^\pm}$ are the resulting pion valence quark densities, 
$v_{Kq}$ and $v_{Ks}$ are the light and strange valon distributions in Kaons, 
and $s_s=\bar{s}_s$ the non-perturbative strange quark distributions in charged pions. 
$P_{KK}$ is the probability density of a Kaon inside a pion and 
$P_{g\pi}(x)=P_{\pi g}(1-x)$ is the charged pion distribution in the 
$\left|\hat{\pi}^\pm_\circ~g\right>$ fluctuation. The hadronic distributions 
inside pions, $P_{\pi g}$ and $P_{KK}$, are given by similar formulas to 
those of eqs.~(\ref{eq3})-(\ref{eq5}). 

It is interesting to note that light quarks in a $\left|KK\right>$ 
fluctuation contribute to the charged pion low $Q^2$ valence densities 
but not to their intrinsic sea distributions.  This is because although there are 
non-perturbative contributions to the light quark distributions in charged pions, 
they appear 
in the $\bar{u}\,(u)$ and $d\,(\bar{d})$ sectors but not in the $u\,(\bar{u})$ 
and $\bar{d}\,(d)$ for the $\pi^-\,(\pi^+)$ respectively \footnote{Recall the flavor
structure of the particles involved: $\pi^+(u\bar d) \rightarrow 
K^+(u\bar s)\bar{K}^0(s\bar d)$ and $\pi^-(\bar ud) \rightarrow 
K^-(\bar u s){K}^0(\bar s d)$.}.
In turn, for neutral pions the hadronic Fock state expansion has the form 
\beq
\left|\pi^0\right> = b_0\left|\pi^0_\circ\right> + b_1 \left|\hat{\pi}_\circ^0\,g\right> 
+ b_2 \left|\pi^-\,\pi^+\right> + \frac{b_3}{\sqrt{2}} \left[ \left|K^-\,K^+\right> 
- \left|K^0\,\bar{K}^0\right> \right] + ... \; .
\label{eq15a}
\eeq

Then, by analogy with eqs.~(\ref{eq14}),  we can define
\bea
v_{q/\pi^0}(x,Q_v^2) &=& 
\frac{1}{2} |b_0|^2\,v_{\Pi}(x) + \frac{1}{2} |b_1|^2\,
\int_x^1{ \frac{dy}{y}P_{g\pi}(y)\,
v_\Pi\left( \frac{x}{y} \right) } \nonumber \\
& &+ \frac{|b_3|^2}{2}\,\int_x^1{ \frac{dy}{y}P_{KK}(y)\,v_{Kq}
\left( \frac{x}{y} \right) } 
\label{eq15b}
\eea
for the valence quark densities at the low $Q_v^2$ scale, and 
\bea
s_{u/\pi^0}(x,Q_v^2) & = & s_{\bar{u}/\pi^0}(x,Q_v^2) = 
s_{d/\pi^0}(x,Q_v^2)  = s_{\bar{d}/\pi^0}(x,Q_v^2) \nonumber \\
& = & |b_2|^2 \,\int_x^1{ \frac{dy}{y}P_{\pi\pi}(y)\,v_\Pi\left( \frac{x}{y} \right) } 
\nonumber \\
s_s(x,Q_v^2) & = & \bar{s}_s(x,Q_v^2) = |b_3|^2\,
\int_x^1{ \frac{dy}{y} P_{KK}(y)\,v_{Ks} \left( \frac{x}{y} \right) }, 
\label{eq15c}
\eea
for the intrinsic up, down and strange seas. Gluon distributions are given by
eq.~(\ref{eq11}) for both neutral and charged pions.

It should be noted that, although considering the $s_{q/\pi^0}$ densities as part of
the intrinsic light sea or part of the valence densities in the $\pi^0$ is a matter of 
convention, the low $Q_v^2$ structure of the $\pi^0$ is different to the structure 
of charged pions. The difference is precisely given by the contribution of the 
$\left|\pi^-\pi^+\right>$  fluctuation which can only occur in a $\pi^0$ state.

As a final result, notice that
the intrinsic quark-antiquark sea of pions turns out to be symmetric as a consequence 
of the hadronic structure of the fluctuations. This is in contrast to the tipically
unequal intrinsic quark and antiquark distributions of the nucleon (see e.g. 
\cite{christiansen} and  Refs. therein).


\section{Conclusions}\label{sec4}

In this paper we have analysed some important consequences of making
a hadronic Fock state expansion of the nucleon and pion low $Q^2$ wave-functions
out of a novel mechanism for generating the cloud. 
We have shown that within such scheme it is possible to generate not only  
non-perturbative quark-antiquark distributions but also the gluon sea needed 
at the low $Q^2$ starting scale for DGLAP evolution.

These non-pertubative quarks and gluons are responsible for the bound nature 
of any hadron state, as they bring about the interactions between valence quarks. 
The non-perturbative quarks and gluons can be consistently identified 
with the so called intrinsic sea, in contrast to the extrinsic sea. On the 
other hand, the extrinsic sea is perturbatively generated by the probe momentum $Q^2$, 
and is part of the own structure of valons, as discussed long time ago 
by Hwa~\cite{hwa2}.

In this sense our approach leads to a unification of two different pictures of the 
hadron structure; namely, the early picture of (non-interacting) valons~\cite{hwa2}, 
and the intrinsic sea idea of Brodsky {\it et al.}~\cite{brodsky}, 
which provides the binding agent for the bound hadron state. On the other hand, 
our approach allows a full representation of the non-perturbative processes 
giving rise to hadronic quantum fluctuations. This fluctuations are due to 
the perturbative production of a $q\bar q$ pair which recombines with the 
remaining valons. Thus a connection between the physics of hadronic reactions 
and that of hadronic fluctuations is established through the well known 
recombination mechanism. 

A remarkable feature of the approach is that neutral pion fluctuations
are here inhibited and, in turn, non-perturbative gluons take place.
The reason is that neutral unflavored structures like the initial
$v_q\bar{q}$ objects, are more likely to recombine rapidly into gluons 
than into neutral pions, in contrast to flavored structures like $v_q\bar{q}\,'$ 
which cannot do so. Thus, the hypothetical cloud of quantum fluctuations like 
$\left|\hat{p}_\circ~\pi^0\right>$ does not contribute to the sum over $\left|B_iM_i\right>$ 
in the RHS of eq.~(\ref{eq1}) but to the second term, $\left|\hat{p}_\circ~g\right>$, 
providing the source of valence-like gluons in the proton. 

Thus, within our scheme not only intrinsic quarks and antiquarks but also gluons are
generated through quantum fluctuations of the low $Q^2$ hadron ground state.

Concerning pions, we have calculated their quark-antiquark and gluon distributions 
at low $Q^2$. We have also shown that the non-perturbative structure of charged 
and neutral pions are different. The difference arises from the 
$\left|\pi^+\pi^-\right>$ fluctuation appearing in the hadronic Fock state 
expansion of the $\pi^0$ wave-function but not in the charged pion ones. 
Finally, we have shown that the pionic intrinsic quark and antiquark distributions 
are symmetric, as a result of the specific features of its quantum fluctuations. 
This is in contrast to the structure of generic baryons which have asymmetric 
intrinsic quark and antiquark distributions~\cite{christiansen, magnin}. 
However, it should be noted that for mesons containing a light and a heavier 
valence quark, the situation is different and the intrinsic quark-antiquark 
sea must be asymmetric~\cite{uniandes}. 

Summarizing, we have proposed a possible scenario for the origin of the 
valence-like sea quark and gluon distributions nedeed at the low (input) 
scale in order to describe the experimental DIS data for nucleons and pions. We have 
also discussed the low scale structure of charged pions and shown that, 
besides valence quarks, the model predicts only gluon and strange intrinsic sea 
distributions as a suitable low $Q^2$ starting point for perturbative DGLAP evolution. 
On the other hand, for neutral pions, intrinsic light quark-antiquark distribution have 
to be considered as well. This signals a remarkable difference between the 
non-perturbative structure of neutral and charged pions.


\section*{Acknowledgments}
We acknowledge R. Vogt for useful comments. 
J.M. is partially supported by COLCIENCIAS, the Colombian Agency for Science 
and Technology, under Contract No. 242-99.

\newpage

\newpage

\newpage

\newpage

\end{document}